\documentclass[12pt]{article}
\usepackage[centertags]{amsmath}
\usepackage{amsfonts}
\usepackage{amssymb}
\usepackage{graphicx}
\usepackage{amssymb}
\def\cs{\mbox{${\rm c}$}}
\def\s{\mbox{${\rm s}$}}
\usepackage[verbose,colorlinks=true,naturalnames=true,linkcolor=blue,]
{hyperref}

\newcounter{rown}

\begin{document}

%% \title{Two-parameter Jordanian Quantum Deformations of\\Lorentz
%% and Poincar\'{e} Algebras for $D=4$ and $3$\\[15pt]%%}
\title{Jordanian Twist Quantization of $D=4$ Lorentz and
Poincar\'{e} Algebras and $D=3$ Contraction Limit
%% \\[15pt]%%}
%% New Twist Quantization of $D=4$ Lorentz and
%% Poincar\'{e} Algebras and $D=3$ Contraction Limit
}

\author{A. Borowiec$^{1)}$, J. Lukierski$^{1)}$ and V.N. Tolstoy$^{1),2)}$
\\ \\
$^{1)}$Institute for Theoretical Physics,
\\University of Wroc{\l}aw, pl. Maxa Borna 9,
\\50--205 Wroc{\l}aw, Poland\\
\\$^{2)}$Institute of Nuclear Physics,
\\Moscow State University, 119 992 Moscow, Russia}

\date{}
\maketitle
\begin{abstract}
We describe in detail two-parameter nonstandard quantum deformation of $D=4$
Lorentz algebra $\mathfrak{o}(3,1)$, linked with Jordanian deformation of
$\mathfrak{sl} (2;\mathbb{C})$. Using twist quantization technique we obtain the
explicit formulae for the deformed coproducts and antipodes. Further extending
the considered deformation to the $D=4$ Poincar\'{e} algebra we obtain a new
Hopf-algebraic deformation of four-dimensional relativistic symmetries with
dimensionless deformation parameter. Finally, we interpret $\mathfrak{o}(3,1)$
as the $D=3$ de-Sitter algebra  and calculate the contraction limit
$R\rightarrow\infty$ ($R$ -- de-Sitter radius) providing explicit
Hopf algebra structure for the quantum deformation of the $D=3$ Poincar\'{e} algebra
(with masslike deformation parameters), which is the two-parameter light-cone
$\kappa$-deformation of the $D=3$ Poincar\'{e} symmetry.
\end{abstract}

\section{Introduction}

In the last decade there were considered quantum field theories on noncommutative
space-time, after it has been shown that the noncommutativity might follow from
quantum gravity corrections (see e.g. \cite{DFR}) or an open string theory with
anti-symmetric tensor field background (see e.g. \cite{SW}). For the simple examples
of noncommutativity, in particular Heisenberg-like space-time commutators
($\theta_{\mu\nu}={\rm const}$) it has been shown \cite{W,CKNT,ABDMSW,LW,GKMG} that
the theory with noncommutative space-time is covariant under twisted Poincar\'{e}
symmetry. In such approach the violation of classical Poincar\'{e} invariance,
e.g., by constant tensor $\theta_{\mu\nu}$, can be equivalently described by
twisting of classical symmetry \cite{LW}. Because twist quantizations of the
Lorentz and Poincar\'{e} algebras are classified by classical $r$-matrices satisfying
homogeneous Yang-Baxter (YB) equations it is interesting to approach the problem
of possible violations of Lorentz and Poincar\'{e} symmetries by considering new,
in particular non-Abelian, twist quantizations\footnote{By non-Abelian twist we mean
a twist two-tensor with a support in a non-Abelian algebra.}.

All quantum deformations of relativistic symmetries are described by Hopf-algebraic
deformations of Lorentz and  Poincar\'{e} algebras. Such quantum deformations 
determine infinitesimally Lorentz and Poincar\'{e} Poisson structures.
These Poisson structures are described by classical $r$-matrices satisfying
homogeneous  as well as inhomogeneous (modified) YB equations
and they were classified already some time ago by S. Zakrzewski (see \cite{Z1}
for the Lorentz classical $r$-matrices and \cite{Z2} for  the Poincar\'{e}
classical matrices). In \cite{Z1} there are provided four classical
$\mathfrak{o}(3,1)$ $r$-matrices and in \cite{Z2} one finds 21 cases describing
different deformations of Poincar\'{e} symmetry, with various numbers of free
parameters.

In this paper we would like to describe the explicit Hopf algebra for an
important nonstandard deformation of the $D=4$ Lorentz algebra generated by
two-parameter Jordanian classical $r$-matrix.

Part of the results presented in this paper has been given in our short report
\cite{BLT1} where the complex algebra basis was used. In contrast to \cite{BLT1}
here all obtained results are represented in real (more physical) basis of the
Lorentz algebra. In this case a twist two-tensor depends on new generators
$\sigma$ and $\varphi$ which are operator analogs of polar coordinates for
a complex plane. Moreover we interpret $D=4$ Lorentz algebra as $D=3$ de-Sitter
algebra and following the quantum contraction method applied firstly to the
$q$-deformed $D=4$ anti-de-Sitter (dS) algebra \cite{LNRT} we calculate the
contraction limit $R\rightarrow\infty$ ($R$ -- de-Sitter radius) providing a
deformation  of $D=3$ Poincar\'{e}
algebra. Subsequently, we present in explicit form the Hopf structure describing two-parameter
extension of the light-cone $\kappa$-deformation for $D=3$ Poincar\'{e} symmetry.

The plan of the paper is the following. In Sect.2 we describe different bases
of the real Lorentz algebra, $\mathfrak{o}(3,1)$, and recall the classification
of all possible quantum deformations of the $D=4$ Lorentz algebra \cite{Z2}.
In Sect.3 we calculate the twist function corresponding to the non-Abelian Jordanian
classical $r$-matrix (\ref{v17}) and describe explicitly  the deformed coproducts
and antipodes in classical $\mathfrak{o}(3,1)$ basis. In Sect.4 we obtain a deformed
$D=4$ Poincar\'{e}-Hopf algebra by adding four momentum generators and calculating
their twisted coproducts. In Sect.5 we rewrite our quantum deformation in terms
of $D=3$ de-Sitter algebra and we perform the contraction, providing two-parameter
light cone $\kappa$-deformation of the $D=3$ Poincar\'{e} symmetry. In Sect.6 we
present possible extensions of the present work.

\setcounter{equation}{0}
\section{$D=4$ Lorentz algebra and its classical $r$-matrices}

The classical canonical basis of the $D=4$ Lorentz algebra, $\mathfrak{o}(3,1)$,
can be described by anti-Hermitian six generators ($h$, $e_{\pm}$, $h'$,
$e'_{\pm}$) satisfying the following nonvanishing commutation relations:
\begin{eqnarray}
&[h,\,e_{\pm}]\;=\;\pm e_{\pm}\,,\qquad [e_{+},\,e_{-}]\;=\;2h~,
\label{v1}
\\[8pt]
&[h,\,e'_{\pm}]\;=\;\pm e'_{\pm}~,\qquad [h',\,e_{\pm}]\;=\;\pm
e'_{\pm}~,\qquad [e_{\pm},\,e'_{\mp}]\;=\;\pm2h'~ ,\label{v2}
\\[8pt]
& [h',\,e'_{\pm}]\;=\;\mp e_{\pm}^{}~,\qquad
[e'_{+},\,e'_{-}]\;=\;-2h~,\label{v3}
\end{eqnarray}
and moreover
\begin{equation}
x^*\;=\;-x\qquad (\forall\;x\;\in\; \mathfrak{o}(3,1))~.\textbf{}
\label{v4}
\end{equation}\texttt{}
The formulas (\ref{v1})--(\ref{v3}) can be rewritten as the following three
$ \mathfrak{o}(3)$-covariant relations describing Lorentz algebra in "physical"
basis ($i,j,k=1,2,3$)\footnote{In what follows the symbol $"\imath"$ means
the imaginary unit, $\imath^2=-1$.}
\begin{equation}
[M_i,\,M_j ]\;=\;\imath\,\epsilon_{ijk}\,M_k~,\quad\;
[M_i,\,N_j]\;=\;\imath\,\epsilon_{ijk}N_k~,\quad\;
[N_i,\,N_j]\;=\;-\imath\,\epsilon_{ijk}M_k~,\label{v5}
\end{equation}
where the three-dimensional rotation generators $M_i$ and boosts $N_i$ are related
%% in \cite{BLT1}
with the canonical basis (\ref{v1})--(\ref{v3}) as follows
\begin{equation}\label{v6}
\begin{array}{rcccl}
h\!\!&=\!\!&\imath\,N_3~,\qquad\quad e_{\pm}\!\!&=\!\!&\imath\,(N_1\pm\,M_2)~,
\\[7pt]
h'\!\!&=\!\!&-\imath\,M_3~,\qquad e'_{\pm}\!\!&=\!\!&\imath\,(\pm N_2-M_1)~,
\end{array}
\end{equation}
and they are Hermitian
\begin{equation}\label{v7}
\begin{array}{rcccl}
M_i^*\!\!&=\!\!&M_i^{}~,\qquad\quad
N_i^*\!\!&=\!\!&N_i^{}\qquad(i=1,2,3)~.
\end{array}
\end{equation}

It should be stressed that the realization (\ref{v6}) of the canonical basis in
terms of the physical generators is not unique. Indeed, it is easy to see that
the formulas (\ref{v5}) are invariant with respect to the cyclic permutation
of the indexes $1\rightarrow2\rightarrow3\rightarrow1$ for the generators $M_i$
and $N_i$. Therefore if we apply such cyclic replacements to the generators
of the right side of ((\ref{v6})) we obtain another physical assignments of the
canonical basis. In order to obtain suitable contraction limit (see Sect.4) we
perform the cyclic replacements two times and we get other physical assignment
of the basis (\ref{v1})--(\ref{v3}):
\begin{equation}\label{v8}
\begin{array}{rcccl}
h\!\!&=\!\!&\imath\,N_2~,\qquad\quad e_{\pm}\!\!&=\!\!&\imath\,(N_3\pm\,M_1)~,
\\[7pt]
h'\!\!&=\!\!&-\imath\,M_2~,\qquad e'_{\pm}\!\!&=\!\!&\imath\,(\pm N_1-M_3)~.
\end{array}
\end{equation}

The complete classification of $D=4$ Lorentz quantum algebras in \cite{Z1} is
provided by the following  list of the corresponding classical $r$-matrices
(see also \cite{WZ}):\\ 1. Standard $r$-matrices related with realification of
Drinfeld-Jimbo deformation of $\mathfrak{sl}(2,\mathbb{C})$ \cite{D1,J,SWZ}
\begin{eqnarray}\label{v9}
r_1(\alpha,\beta,\gamma)\!\!&=\!\!&\alpha(e'_{+}\wedge e_{-}+
e_{+}\wedge e'_{-})\,+\,\beta\,(e_{+}\wedge e_{-}- e'_{+}\wedge
e'_{-})+\gamma h\wedge h'~,
\\[5pt]\label{v10}
r_2(\alpha)\!\!&=\!\!&\alpha\Bigl(e'_{+}\wedge e_{-}+ e_{+}\wedge
e'_{-}+\frac{1}{2}h\wedge h'\Bigr)\,\pm\, e_{+}\wedge e'_{+}~.
\end{eqnarray}
2. Nonstandard $r$-matrices, related with realification of Jordanian deformation
of $\mathfrak{sl}(2,\mathbb{C})$ \cite{Oh,Og,ACCS}
\begin{eqnarray}\label{v11}
r_3(\alpha,\beta)\!\!&=\!\!&\alpha(h\wedge e_{+}-h'\wedge
e'_{+})\,+\,\beta\,e_{+}\wedge e'_{+}~,
\\[5pt]\label{v12}
r_4(\alpha)\!\!&=\!\!&\alpha\,h\wedge e_{+}~.
\end{eqnarray}
Further we shall assume that the $r$-matrices are anti-Hermitian, i.e.
$r_i^*=-r_i$, therefore all the parameters in (\ref{v9})--(\ref{v12}) are
purely imaginary.

The standard quantum deformations (\ref{v9}) and (\ref{v10}) do satisfy modified
YB equation and can not be extended to Poincar\'{e} algebra \cite{M, PW}.
The nonstandard quantum deformations (\ref{v11}) and (\ref{v12}) can be extended
to the whole Poincar\'{e} algebra and provide new deformation of
relativistic symmetries. Because the quantization of (\ref{v12}) by means of the
Ogievetsky twist (see \cite{Og}) is straightforward and describes the well-known
Jordanian deformation, we shall consider here more in detail only the
two-parameter deformation generated by the $r$-matrix (\ref{v11}).

Further we shall  employ as well the complex basis of Lorentz algebra
$(\mathfrak{o}(3,1)\simeq\mathfrak{o}(3;\mathbb{C})\oplus\mathfrak{\overline{o}}
(3,\mathbb{C}))$described by two commuting sets of complex generators:
\begin{eqnarray}\label{v13}
H_1\!\!&=\!\!& \frac{1}{2}\,(h+\imath h')~,\qquad
E_{1\pm}\;=\;\frac{1}{{2}}\,(e_{\pm}^{}+ \imath e'_{\pm})~,
\\[5pt]\label{v14}
H_2\!\!&=\!\!& \frac{1}{{2}}\, (h-\imath h')~,\qquad
E_{2\pm}\;=\;\frac{1}{{2}}\,(e_{\pm}^{}-\imath e'_{\pm})~,
\end{eqnarray}
which satisfy the relations (compare with (\ref{v1}))
\begin{equation}\label{v15}
[H_i,\,E_{i\pm}]\;=\;\pm E_{i\pm}~,\qquad [E_{i+},\,E_{i-}]\;=\;2
H_i\qquad(i=1,2)~,
\end{equation}
where the sets $(H_1,\,E_{1\pm})$ and $(H_2,\,E_{2\pm})$ do commute mutually.
The $*$-operation describing the real structure acts on the generators $H_i$, and
$E_{i\pm}$ ($i=1,2$)  as follows
\begin{equation}\label{v16}
H_1^*\;=\;-H_2^{}~,\qquad E_{1\pm}^*\;=\;- E_{2\pm}^{}~,\qquad
H_2^*\;=\;-H_1^{}~,\qquad E_{2\pm}^*\;=\;-
E_{1\pm}^{}~.
\end{equation}

The classical $r$-matrix $r_3(\alpha,\beta)$ (see (\ref{v11})) in complex
basis (\ref{v13}), (\ref{v14}) takes the form
\begin{eqnarray}\label{v17}
&r_3^{}(\alpha,\beta)\;=\;r'_3(\alpha)+r''_3(\beta)~,
\\[7pt]\label{v18}
&r'_3(\alpha)\;:=\;2\alpha\,(H_1\wedge E_{1+} + H_2\wedge
E_{2+})~,\qquad r''_3(\beta)\;:=\;2i\beta\,E_{1+}\wedge E_{2+}~.
\end{eqnarray}

\setcounter{equation}{0}
\section{Two-parameter nonstandard deformation of $\mathfrak{o}(3,1)$}

The first classical $r$-matrix $r'_3(\alpha)$ in (\ref{v17}) is a sum of two
Jordanian classical $r$-matrices $\alpha H_1\wedge E_1$ and $\alpha H_2\wedge E_2$
which mutually commute. From this property follows that the first twisting
two-tensor $F'$ corresponding to the  Jordanian type $r$-matrix $r'_3(\alpha)$
is a product of two Jordanian twists with the same deformation parameter $\alpha$:
\begin{equation}\label{jt1}
F'\;:=\;F_{1J}^{}\,F_{2J}^{}\;=\; F_{2J}^{}\,F_{1J}^{}\;=\;
\exp{(H_1\otimes\sigma_1+H_2\otimes\sigma_2)}~,
\end{equation}
where
\begin{equation}\label{jt2}
\sigma_i\,=\,\ln{(1+2\alpha E_{i+})}~,\qquad i=1,2~.
\end{equation}
It should be noted that the Jordanian two-tensor (\ref{jt1}) is $*$-unitary, i.e.
\begin{equation}\label{jt3}
F'{}^*\;=\;F'{}^{-1}~.
\end{equation}
Since after twisting by the two-tensor (\ref{jt1}) the generators (\ref{jt2})
have the primitive co-products
\begin{equation}\label{jt4}
\Delta^{(F')}(\sigma_i)\;=\;\sigma_i\otimes1+1\otimes\sigma_i~,\qquad i=1,2~,
\end{equation}
therefore the two-tensor\footnote{In the limit $\alpha\rightarrow0$ the two-tensor
$F''$  goes to $\exp{(2i\beta\,E_{1+}\wedge E_{2+})}$ which is the twist in the
direction of $r''_3(\alpha=0,\beta)$ (see (\ref{v11})).}$^,$\footnote{It should
be mentioned that modulo the explicit parameters dependence the algebraic
form of the twist (\ref{jt5}) was early pointed by Kulish and  Mudrov
(see \cite{KM}), item 3 on the list on p.6)}
\begin{equation}\label{jt5}
F''\;=\;\exp{\Bigl(\imath\frac{\beta}{2\alpha^2}\,\sigma_1\wedge\sigma_2\Bigr)}
\end{equation}
satisfies the cocycle condition (see \cite{D2})
\begin{equation}\label{jt6}
F''{}^{12}(\Delta^{(F')}\otimes{\rm id})(F'')\;=\; F''{}^{23}({\rm
id}\otimes\Delta^{(F')})(F'')~,
\end{equation}
and the "unital" normalization condition
\begin{equation}\label{jt7}
(\epsilon \otimes{\rm id})(F'')\;=\;({\rm id}\otimes\epsilon)(F'')=1~.
\end{equation}
Thus the complete twisting two-tensor $F(\alpha,\beta)$ corresponding to the
Jordanian type $r$-matrix (\ref{v17}) is given as follows %%\footnote{It should
%% be mentioned that modulo the explicit parameters dependence the algebraic
%% form of the twist (\ref{jt5}) was firstly pointed by Kulish and  Mudrov
%% (see \cite{KM}), item 3 on the list on p.6)}
\begin{equation}\label{jt8}
F\;:=\;F(\alpha,\beta)\;=\;F''F'\;=\;
\exp{\Bigl(\imath\frac{\beta}{2\alpha^2}\;\sigma_1\wedge\sigma_2\Bigr)}\,
\exp{(H_1\otimes\sigma_1+H_2\otimes\sigma_2)}~.
\end{equation}

Let us express this function in terms of the generators ($h$, $e_{\pm}$, $h'$,
$e'_{\pm})$. Firstly we notice that the generators $\sigma_i$ given by (\ref{jt2})
do not change after the second twist $F''$, i.e. they have the primitive
coproducts
\begin{equation}\label{jt9}
\Delta^{(F)}(\sigma_i)\;=\;F''\Delta^{(F')}(\sigma_i)
F''{}^{-1}\;=\;\sigma_i\otimes1+1\otimes\sigma_i~,\qquad
i=1,2~.
\end{equation}
Therefore if instead of $\sigma_1$ and $\sigma_2$ we introduce the new
generators $\sigma$ and $\varphi$ by the formulas
\begin{equation}\label{jt10}
\sigma_1\;=\;\sigma+\imath\,\varphi~,\qquad\sigma_2\;=\;\sigma-\imath\,\varphi~,
\end{equation}
where
\begin{eqnarray}\label{jt11}
\sigma\!\!& =\!\!&\frac{1}{2}\ln\left[(1+\alpha\,e_+)^2+
(\alpha\,e'_+)^2 \right]~,\qquad \varphi\;=
\;\arctan{\frac{\alpha\, e'_+}{1+\alpha\,e_+}}~,
\end{eqnarray}
then the new generators also have the primitive coproducts
\begin{equation}\label{jt12}
\Delta^{(F)}(\sigma)\;=\;\sigma\otimes1+
1\otimes\sigma~,\qquad\Delta^{(F)}(\varphi)\;=\;\varphi\otimes1+
1\otimes\varphi~,
\end{equation}
and moreover they are $*$-Hermitian
\begin{equation}\label{jt13}
\sigma^*\;=\;\sigma~,\qquad \varphi^*\;=\;\varphi~.
\end{equation}
The formulas inverse to (\ref{jt11}) have the form
\begin{eqnarray}\label{jt14}
\alpha\,e_{+}\!\!&=\!\!&e^{\sigma}\cos{\varphi}-1~,\qquad
\alpha\,e'_{+}\;= \;e^{\sigma}\sin{\varphi}~.
\end{eqnarray}
Substituting (\ref{jt10}) and (\ref{v13}), (\ref{v14}) in (\ref{jt8}) we
obtain the following formula for the twist two-tensor in terms of the
canonical $\mathfrak{o}(3,1)$-basis (\ref{v1})--(\ref{v3}):
\begin{equation}\label{jt15}
F(\alpha,\beta)\;=\;
\exp{\Bigl(\frac{\beta}{\alpha^2}\;\sigma\wedge\varphi\Bigr)}\,
\exp{(h\otimes\sigma-h'\otimes\varphi)}~.
\end{equation}
This two-tensor is $*$-unitary, i.e.
\begin{equation}\label{jt16}
F^*(\alpha,\beta)\;=\;F^{-1}(\alpha,\beta)~.
\end{equation}

Deformed coproducts for the canonical $\mathfrak{o}(3,1)$-basis
can be obtain in two ways. First way is to apply the twisting two-tensor in
the form (\ref{jt8}) to the trivial coproducts of the complex generators $H_i$,
$E_i$, $F_i$ ($i=1,2$) and then by using the formulas (\ref{v13}) and (\ref{v14})
to derive the coproducts and antipodes for the canonical basis. Other way is to
apply the twisting two-tensor in the form (\ref{jt15}) directly to the trivial
coproducts of the canonical generators (\ref{v1})--(\ref{v2}),
($\Delta_{\alpha,\beta}(\,\cdot\,):=F\Delta(\,\cdot\,)F^{-1}$). For convenience we
shall use the notations: $\cs_{\varphi}:=\cos\varphi$ and $\s_{\varphi}:=\sin\varphi$.
We obtain the following formulae:
\begin{equation}\label{jt17}
\begin{array}{rcl}
\Delta_{\alpha,\beta}(h)\!\!&=\!\!&\displaystyle h\otimes e^{-\sigma}
\cs_{\varphi}+1\otimes h+h'\otimes e^{-\sigma}\s_{\varphi}+\frac{\beta}{\alpha^2}
\bigl(e^{-\sigma}\s_{\varphi}\otimes e^{-\sigma}(\sigma\cs_{\varphi}-
\varphi\s_{\varphi})
\\[8pt]
\!\!&\!\!&+(e^{-\sigma}\cs_{\varphi}-1)\otimes e^{-\sigma}(\varphi
\cs_{\varphi}+\sigma\s_{\varphi})-\sigma\otimes e^{-\sigma}\s_{\varphi}-
\varphi\otimes(e^{-\sigma}\cs_{\varphi}-1)\bigl),
\end{array}
\end{equation}
\begin{equation}\label{jt18}
\begin{array}{rcl}
\Delta_{\alpha,\beta}(h')\!\!&=\!\!&\displaystyle h'\otimes e^{-\sigma}
\cs_{\varphi}+1\otimes h'-h\otimes e^{-\sigma}\s_{\varphi}+\frac{\beta}{\alpha^2}
\bigl(-e^{-\sigma}\s_{\varphi}\otimes e^{-\sigma}(\sigma\s_{\varphi}+
\varphi\cs_{\varphi})
\\[8pt]
\!\!&\!\!&+(e^{-\sigma}\cs_{\varphi}-1)\otimes e^{-\sigma}(\sigma
\cs_{\varphi}-\varphi\s_{\varphi})+\varphi\otimes e^{-\sigma}\s_{\varphi}-
\sigma\otimes(e^{-\sigma}\cs_{\varphi}-1)\bigl),
\end{array}
\end{equation}
The remaining coproducts $\Delta_{\alpha,\beta}(e_{-})$ and
$\Delta_{\alpha,\beta}(e'_{-})$ are given by lengthy formulae, and therefore we
present the $\beta$-dependent terms using the complex basis
\begin{equation}\label{jt19}
\begin{array}{rcl}
&&\Delta_{\alpha,\beta}(e_{-})\,=\,e_{-}\otimes e^{-\sigma}
\cs_{\varphi}+1\otimes e_{-}+e_{-}'\otimes e^{-\sigma}\s_{\varphi}
\\[8pt]
\!\!&\!\!&\quad+\,\alpha\Bigl(h\otimes\bigl(\{h',e^{-\sigma}\cs_{\varphi}\}+
\{h',e^{-\sigma}\s_{\varphi}\}\bigr)-h'\otimes\bigl(\{h',e^{-\sigma}\cs_{\varphi}\}-
\{h,e^{-\sigma}\s_{\varphi}\}\bigr)
\\[8pt]
\!\!&\!\!&
\quad+\,(h^{2}-{h'}^{2})\otimes(e^{-2\sigma}\cs_{2\varphi}-
e^{-\sigma}\cs_{\varphi})+2hh'\otimes(e^{-2\sigma}\s_{2\varphi}-
e^{-\sigma}\s_{\varphi})\Bigl)
\\[4pt]
\!\!&\!\!&\quad+\;\displaystyle\frac{\imath\beta}{4\alpha}\,
\bigl(\mathcal{E}_{1}+\mathcal{E}_{1}^*\bigr)-\frac{\beta^{2}}{4\alpha^{3}}\,
\bigr(\mathcal{E}_{2}+\mathcal{E}_{2}^*\bigr)~,
\end{array}
\end{equation}
\begin{equation}\label{jt20}
\begin{array}{rcl}
&&\Delta_{\alpha,\beta}(e_{-}')\,=\,\displaystyle e_{-}'\otimes e^{-\sigma}
\cs_{\varphi}+1\otimes e_{-}'+e_{-}\otimes e^{-\sigma}\s_{\varphi}
\\[8pt]
\!\!&\!\!&\quad+\,\alpha\Bigl(h'\otimes\bigl(\{h,e^{-\sigma}\cs_{\varphi}\}+
\{h',e^{-\sigma}\s_{\varphi}\}\bigr)+h\otimes\bigl(\{h',e^{-\sigma}\cs_{\varphi}\}-
\{h,e^{-\sigma}\s_{\varphi}\}\bigl)
\\[8pt]
\!\!&\!\!&
\quad-\,(h^{2}-{h'}^{2})\otimes(e^{-2\sigma}\s_{2\varphi}-e^{-\sigma}\s_{\varphi})
+2hh'\otimes(e^{-2\sigma}\cs_{2\varphi}-e^{-\sigma}\cs_{\varphi})\Bigl)
\\[4pt]
\!\!&\!\!&\quad+\;\displaystyle\frac{\beta}{4\alpha}\,\bigl(\mathcal{E}_{1}-
\mathcal{E}_{1}^*\bigr)+\frac{\imath\beta^{2}}{4\alpha^{3}}\,
\bigr(\mathcal{E}_{2}-\mathcal{E}_{2}^*\bigr)~,
\end{array}
\end{equation}
where
\begin{equation}\label{jt21}
\begin{array}{rcl}
\mathcal{E}_{1}\!\!&=\!\!&\{\tilde{H}_{1}^{},e^{-\sigma_{1}}\}\otimes\sigma_{2}
e^{-\sigma_{1}}-\sigma_{2}\otimes\{\tilde{H}_{1}^{},e^{-\sigma_{1}}\}+\Lambda_{1}^{}
\otimes\{\tilde{H}_{1}^{},e^{-\sigma_{1}}\}\sigma_{2}
\\[8pt]
&&
\,-\,\{\tilde{H}_{1}^{},\Lambda_{1}^{}\}\otimes\sigma_{2}\Lambda_{1}^{}
e^{-\sigma_{1}}-\{\tilde{H}_{1}^{},\sigma_{2}\}\otimes\Lambda_{1}^{}e^{-\sigma_{1}},
\end{array}
\end{equation}
\begin{equation}\label{jt22}
\mathcal{E}_{2}\;=\;\sigma_{2}^{2}\otimes\Lambda_{1}e^{-\sigma_{1}}+
\Lambda_{1}e^{-\sigma_{1}}\otimes\sigma_{2}^{2}e^{-\sigma_{1}}+
\Lambda_{1}^{2}\otimes\sigma_{2}^{2}\Lambda_{1}e^{-\sigma_{1}}-
2\Lambda_{1}\sigma_{2}\otimes\sigma_{2}\Lambda_{1}e^{-\sigma_{1}},
\end{equation}
and
\begin{equation}\label{jt23}
\tilde{H}_{1}^{}\;=\;h+\imath h'~,\quad \Lambda_{1}^{}=e^{-\sigma_{1}}-1~,\quad
\sigma_{1}\,=\sigma+\imath\varphi~,\quad\sigma_{2}\,=\sigma-\imath\varphi~.
\end{equation}
Here the brackets $\{\cdot,\cdot\}$ mean the anti-commutator $\{a,b\}=ab+ba$.

Explicit formulae for the antipodes are given by
\begin{equation}\label{jt24}
S_{\alpha,\beta}(\sigma)\;=\;-\sigma~,\qquad
S_{\alpha,\beta}(\varphi)\;=\;-\varphi~,
\end{equation}
\begin{equation}\label{jt25}
S_{\alpha,\beta}(h)=-h\,e^{\sigma}\cs_{\varphi}+h'\,e^{\sigma}\s_{\varphi}~,\quad
S_{\alpha,\beta}(h')=-h'\,e^{\sigma}\cs_{\varphi}-h\,e^{\sigma}\s_{\varphi}~,
\end{equation}
\begin{equation}\label{jt26}
\begin{array}{rcl}
S_{\alpha,\beta}(e_{-})\!\!&=\!\!&-\,e_{-}\,e^{\sigma}\cs_{\varphi}+
e_{-}'\,e^{\sigma}\s_{\varphi}+\alpha(h^{2}-{h'}^{2})(e^{2\sigma}\cs_{2\varphi}+
e^{\sigma}\cs_{\varphi})\quad
\\[7pt]
&&-\;2\alpha hh'(e^{2\sigma}\s_{2\varphi}+e^{\sigma}\s_{\varphi})~,
\end{array}
\end{equation}
\begin{equation}\label{jt27}
\begin{array}{rcl}
S_{\alpha,\beta}(e_{-}')\!\!&=\!\!&-\,e_{-}'\,e^{\sigma}\cs_{\varphi}-
e_{-}\,e^{\sigma}\s_{\varphi}+\alpha(h^{2}-{h'}^{2})(e^{2\sigma}\s_{2\varphi}+
e^{\sigma}\s_{\varphi})\quad
\\[7pt]
&&+\;2\alpha hh'(e^{2\sigma}\cs_{2\varphi}+e^{\sigma}\cs_{\varphi})~.
\end{array}
\end{equation}

Using the relations (\ref{v6}) the formulae (\ref{jt17})--(\ref{jt27}) can be
rewritten in terms of physical generators $M_{j}$ and $N_{j}$ $(j=1,2,3)$.

\setcounter{equation}{0}
\section{Extension of the deformation to Poincar\'{e} algebra}

The $D=4$ Lorentz algebra (\ref{v6}) can be extended to $D=4$ Poincar\'{e}
algebra by adding the mutually commuting four-momentum operators
($P_{0},\,P_{1},\,P_{2},\,P_{3})$ satisfying the relations ($j,k,l=1,2,3$)
\begin{equation}\label{dp1}
\begin{array}{rcl}
[M_{i},\,P_{j}]\!\!& =\!\!&\imath\,\epsilon_{ijk}\,P_{k}~,\qquad\;
[M_{i},\,P_{0}]\;=\;0~,
\\[7pt]
[N_{i},\,P_{j}]\!\!&=\!\!&-\imath\,\delta_{ij}\,P_{0}~,\qquad
[N_{i},\,P_{0}]\;=\;-\imath\,P_{i}~.
\end{array}
\end{equation}
The formulas (\ref{dp1}) say that a linear span of the four-momentum generators
$P_{\mu}$, $(\mu=0,1,2,3)$ is a four-dimensional $\mathfrak{o}(3,1)$-module
with respect to the adjoint action of Lorentz algebra on the four-momentum space.
In order to describe in this space the actions of the canonical basis
$(h,\,e_{\pm},\,h',\,e'_{\pm})$ it is useful to use as a technical
tool (see also \cite{Z2}) the matrix realization of the four-momenta. The matrix
realization is constructed as follows.

First we introduce the following basis in the four-momentum space
\begin{equation}\label{dp2}
\begin{array}{rcl}
P_{11}\!\!&=\!\!&\displaystyle
\frac{1}{2}\Bigl(P_{0}+P_{3}\Bigr)~,\qquad\;\;
P_{12}\;=\;\frac{1}{2}\Bigl(P_{1}-\imath P_{2}\Bigr)~,
\\[7pt]
P_{21}\!\!&=\!\!&\displaystyle
\frac{1}{2}\Bigl(P_{1}+\imath P_{2}\Bigr)~,\qquad\;
P_{22}\;=\;\frac{1}{2}\Bigl(P_{0}-P_{3}\Bigr)~.
\end{array}
\end{equation}
Let $\hat{P}_{ij}$, $(i,j=1,2)$ be the $2\!\times\!2$-matrices
$(\hat{P}_{ij})_{kl}=\delta_{ik}\delta_{jl}$. The matrix realization of the
four-momenta is given by $P_{ij}\;\rightarrow\;\hat{P}_{ij}$.
The arbitrary four-vector $\mathbf{P}:=\sum\limits_{i,j=1,2}\xi_{ij}P_{ij}$,
($\xi_{ij}\in \mathbb{R}$), is represented by the general $2\!\times\!2$-matrix
\begin{equation}\label{dp4}
\mathbf{P}\;\rightarrow\;\hat{\mathbf{P}}\;:=
\;\sum_{i,j=1,2}\xi_{ij}\hat{P}_{ij}\;=\;
\begin{pmatrix}
\xi_{11}&\xi_{12}\\
\xi_{21}&\xi_{22}
\end{pmatrix}~,
\end{equation}
which will be used below to describe the transformation  properties of the generators
(\ref{dp2}) in compact way. The corresponding $2\!\times\!2$-matrix realization
of the Lorentz algebra (\ref{v1})--(\ref{v3}) will be used
\begin{equation}\label{dp5}
h\,\rightarrow\,H\,=\,\frac{1}{2}
\begin{pmatrix}
 1 & \;\;0 \\
 0 & -1
\end{pmatrix},\quad
e_+\,\rightarrow\,E_+\,=\,
\begin{pmatrix}
 0 & 1 \\
 0 & 0
\end{pmatrix},\quad
e_-\,\rightarrow\,E_-\,=\,
\begin{pmatrix}
 0 & 0 \\
 1 & 0
\end{pmatrix},
\end{equation}
\begin{equation}\label{dp6}
h'\,\rightarrow\,H'\,=\,\imath\,H~,\qquad
e'_{+}\,\rightarrow\,E'_{+}\,=\,\imath\,E_{+}~,\qquad
e'_{-}\,\rightarrow\,E'_{-}\,=\,\imath\,E_{-}~.
\end{equation}
The action of any element $x\,\in\,(h,\,e_{\pm},\,h',\, e'_{\pm})$ on the
matrix-vector $\hat{\mathbf{P}}$ is given by (see \cite{Z2}):
\begin{equation}\label{dp7}
[x,\mathbf{P}]\,\rightarrow\,x\triangleright\hat{\mathbf{P}}
\,:=\,X\hat{\mathbf{P}}+\hat{\mathbf{P}}X^{+}~.
\end{equation}
where $X^{+}$ is the hermitian conjugate of $X$. Using the formula (\ref{dp7}) we find,
for example,
\begin{equation}\label{dp8}
h\triangleright\hat{\mathbf{P}}\,=\,
\begin{pmatrix}\xi_{11}&0 \\ 0\;\; &-\xi_{22}\end{pmatrix},\qquad
h'\triangleright\hat{\mathbf{P}}\,=\,\imath
\begin{pmatrix}0&\xi_{21}\\-\xi_{12}&\!\!0\end{pmatrix},
\end{equation}
\begin{equation}\label{dp9}
e_{+}\triangleright\hat{\mathbf{P}}\,=\,
\begin{pmatrix}\xi_{21}+\xi_{12}&\!\xi_{22}\\ \xi_{22} &\!\!\!\!0\end{pmatrix},
\qquad e'_{+}\triangleright\hat{\mathbf{P}}\,=\,\imath
\begin{pmatrix}\xi_{21}-\xi_{12}&\!\xi_{22}\\-\xi_{22}&\!\!\!\!0\end{pmatrix}.
\end{equation}

The classical $r$-matrix (\ref{v10}) for $D=4$ Lorentz algebra satisfies classical
YB equation and provides also the deformation of $D=4$ Poincar\'{e} algebra. The
modification of classical Hopf algebra structure is described by the twist
function (\ref{jt15}). Let us calculate the twisted coproducts of the
four-momenta in our $2\times2$-matrix realization (\ref{dp4}) using the
restricted twist two-tensor $F'(\alpha):=F(\alpha,\beta=0)$. We have
\begin{equation}\label{dp10}
\Delta_{\alpha}(\hat{\mathbf{P}})
\;=\;F'(\alpha)\,\hat{\mathbf{P}}\otimes1\,F'{}^{-1}(\alpha)+
F'(\alpha)\,1\otimes\hat{\mathbf{P}}\,F'{}^{-1}(\alpha)~.
\end{equation}
By employing the formulas (\ref{dp7})--(\ref{dp9}) we find
\begin{equation}\label{dp11}
\begin{array}{rcl}
\Delta_{\alpha}(\hat{\mathbf{P}})\!\!&=\!\!&
\begin{pmatrix}
\xi_{11}\otimes e^{\sigma}&\!\xi_{12}\otimes e^{\imath\varphi}\\[3pt]
\xi_{21}\otimes e^{-\imath\varphi}&\!\xi_{22}\otimes e^{-\sigma}\end{pmatrix}+
1\otimes\begin{pmatrix}\xi_{11}&\!\xi_{12}\\[3pt]\xi_{21}&\!\xi_{22}
\end{pmatrix}
\\[20pt]
&&\quad+\;\,\alpha h\otimes\Biggl\{e^{-\sigma+\imath\varphi}
\begin{pmatrix}\xi_{21}&\xi_{22}\\[3pt]0&0\end{pmatrix}
+e^{-\sigma-\imath\varphi}\begin{pmatrix}\xi_{12}&0\\[3pt]\xi_{22}&0
\end{pmatrix}\!\Biggl\}
\\[20pt]
&&\quad-\;\imath\alpha h'\!\otimes\Biggl\{e^{-\sigma+\imath\varphi}
\begin{pmatrix}\xi_{21}&\xi_{22}\\[3pt]0&0\end{pmatrix}
-e^{-\sigma-\imath\varphi}\begin{pmatrix}\xi_{12}&0\\[3pt]\xi_{22}&0
\end{pmatrix}\!\Biggl\}
\\[20pt]
&&\quad+\;\alpha^{2}(h^{2}+{h'}^{2})\otimes e^{-2\sigma}
\begin{pmatrix}
\xi_{22}&0\\[3pt] 0&0
\end{pmatrix}.
\end{array}
\end{equation}
Because $\partial\hat{\mathbf{P}}/\partial\xi_{ij}=\hat{{P}}_{ij}$, by
differentiating (\ref{dp11}) we obtain the formulas for the twisted coproducts
of the four-momentum components $\hat{P}_{ij}$. These formulas after replacing
the realizations $\hat{P}_{ij}$ by the initial generators $P_{ij}$ take the
following compact form:
\begin{equation}
\begin{array}{rcl}
\begin{pmatrix}
\Delta_{\alpha}(P_{11})&\Delta_{\alpha}(P_{12}) \\[3pt]
\Delta_{\alpha}(P_{21})& \Delta_{\alpha}(P_{22})
\end{pmatrix}\!\!&=\!\!&
\begin{pmatrix}
P_{11}\otimes e^{\sigma}&\!P_{12}\otimes e^{-\imath\varphi} \\[3pt]
P_{21}\otimes e^{\imath\varphi}&\!P_{22}\otimes e^{-\sigma}
\end{pmatrix}+ 1\otimes
\begin{pmatrix}
P_{11}&P_{12}\\[3pt] P_{21}&\!P_{22}
\end{pmatrix}
\\[20pt]
&&\;+\;\alpha h\otimes %% e^{-\sigma}
\begin{pmatrix}
0&\;e^{-\sigma-\imath\varphi}P_{11}\\[3pt]
e^{-\sigma+\imath\varphi}P_{11}\;&\;e^{-\sigma+\imath\varphi}P_{12}+
e^{-\sigma-\imath\varphi}P_{21}
\end{pmatrix}
\\[20pt]
&&\;-\,\imath\alpha h'\otimes %% e^{-\sigma}
\begin{pmatrix}0&\;-e^{-\sigma-\imath\varphi}P_{11}\\[3pt]
e^{-\sigma+\imath\varphi}P_{11}&e^{-\sigma+\imath\varphi}P_{12}-
e^{-\sigma-\imath\varphi}P_{21}
\end{pmatrix}
\\[20pt]
&&\;+\;\alpha^{2}(h^{2}+{h'}^{2})\otimes
\begin{pmatrix}
0&0\\[3pt] 0&e^{-2\sigma}P_{11}
\end{pmatrix}.
\end{array}\label{dp12}
\end{equation}
In analogous way we can calculate the complete twisted coproducts of the
four-momenta using the general twist two-tensor $F(\alpha,\beta)$, but these
formulae are lengthly.

\setcounter{equation}{0}
\section{Two-parameter light-cone deformation of the $D=3$ Poincar\'{e} algebra}

The $D=4$ Lorentz algebra $\mathfrak{o}(3,1)$ (see (\ref{v5})) can be reinterpreted
as describing the $D=3$ de-Sitter algebra where the generators $N_1$, $N_2$
describe two $D=3$ boosts,  $M_{3}$ generates $\mathfrak{o}(2)$
rotations and three curved translations are generated by $M_1$, $M_2$ and $N_3$.
Therefore the formulae (\ref{jt17})--(\ref{jt25}) can be used for the description
of the quantum Hopf-algebraic deformations of $D=3$ de-Sitter algebra. Below
we shall use the formulae (\ref{v8}) relating the physical Lorentz generators
with the mathematical basis (\ref{v1})--(\ref{v3}). In this realization the
classical $r$-matrix (\ref{v11}) has the form
\begin{equation}\label{pl1}
r_{3}(\alpha,\beta)\,=-\alpha\,\bigl(N_{2}\wedge(N_{3}+M_{1})+
M_2\wedge(N_{1}-M_{3})\bigr)-\beta\,(N_{3}+M_{1})\wedge(N_{1}-M_{3}).
\end{equation}
Introducing the standard dS-rescaling $M_1=R\mathcal{P}_2$,
$M_2=-R\mathcal{P}_1$, $N_3=R\mathcal{P}_0$ we obtain for the $r$-matrix
(\ref{pl1}) the following form
\begin{equation}\label{pl2}
r_{3}(\alpha,\beta)\,=-\alpha R\bigl(N_{2}\wedge(\mathcal{P}_{0}+\mathcal{P}_{2})
+\mathcal{P}_{1}\wedge(M_{3}-N_{1})\bigr)-\beta R(\mathcal{P}_{0}+\mathcal{P}_{2})
\wedge\mathcal(M_{3}-N_{1})~.
\end{equation}
Now we put $\alpha=-\imath/\kappa R$, $\beta=-\imath/\kappa' R$ (where $\kappa$
and $\kappa'$ are real masslike parameters) and perform the limit $R\to\infty$.
In such a way we obtain the classical $r$-matrix for the $D=3$ Poincar\'{e}
algebra $(\lim\limits_{R\to \infty}{\cal P}_{\mu}^{}=P_\mu^{}$, $\mu=0,1,2)$:
\begin{equation}
\begin{array}{rcl}\label{pl3}
r_{\!\kappa,\kappa'}\!\!& :=\!\!&\displaystyle\lim\limits_{R\to\infty}
r_{\!3}^{}\Bigl(\frac{-\imath}{\kappa R},\frac{-\imath}{\kappa'R}\Bigr)
\\[12pt]
\!\!&=\!\!&\displaystyle\frac{\imath}{\kappa}\bigl(N_{2}\wedge(P_{0}+P_{2})+
P_{1}\wedge(M_{3}-N_{1})\bigr)+\frac{\imath}{\kappa'}(P_{0}+P_{2})
\wedge(M_{3}-N_{1})~,
\end{array}
\end{equation}
where the parameters $\kappa$ and $\kappa'$ are new dimensionfull deformation
parameters. The $D=3$ Poincar\'{e} algebra is described by the generators $M_{3}$,
$N_{1}$, $N_{2}$ of the $D=3$ Lorentz algebra $\mathfrak{o}(2,1)$, and
the generators $P_{0},\,P_{1},\,P_{2}$ span the threemomentum sector.

The classical $r$-matrix (\ref{pl3}) describes two-parameter light cone
$\kappa$-deformation of the $D=3$ Poincar\'{e} algebra\footnote{For analogous
considerations for the $D=4$ case see \cite{BLT2}.}.
Using the commutation relations for the generators $M_{3}$, $N_{1}$, $N_{2}$,
$P_{\pm}:=P_{0}\pm P_{2}$ it easy to check that the $r$-matrix
$r_{\!\kappa,\kappa'}$, when $\kappa'$ goes to $\infty$, is of Jordanian
type\footnote{Compare with \cite{T}, formulae (2.12), after assignment
$h_{\theta^{}}=iN_{2}$, $e_{\theta}=P_{+}$, $e_{\gamma_{1}^{}}=P_{1}$,
$e_{\gamma_{-1}}=(M_{3}-N_{1})$ (see also \cite{BLT2}).}.
Therefore using known general formulae (see \cite{T,KLM}) we can immediately
write down the twisting two-tensor corresponding to the $r$-matrix
$r_{\!\kappa,\kappa'}$. However we
can obtain also a twisting two-tensor corresponding to the classical $r$-matrix
(\ref{pl3}) by applying the dS contraction  limit to the full twisting two-tensor
(\ref{jt15}). First of all it is easy to obtain in the limit $R\to\infty$ the following formulas:
\begin{equation}\label{pl4}
\sigma_{+}\,:=\,\lim\limits_{R\to\infty}\sigma\,=\,\lim\limits_{R\to\infty}
\frac{1}{2}\ln\Bigl[\Bigl(1+\frac{1}{\kappa}\mathcal{P}_{+}\Bigr)^{2}+
\frac{1}{\kappa^{2}R^2}\bigl(N_{1}+M_{3}\bigr)^{2}\Bigr]\,=
\,\ln\Bigr(1+\frac{1}{\kappa}P_{+}\Bigr)~,
\end{equation}
\begin{equation}\label{pl5}
\lim\limits_{R\to\infty}\,(R\varphi)\,=\,\lim\limits_{R\to\infty}
R\cdot\arctan\frac{(N_{1}-M_{3})/\kappa R}{(1+\mathcal{P}_{+}/\kappa)}\,=\,
\frac{1}{\kappa}(N_{1}-M_{3})e^{-\sigma_{+}}.
\end{equation}
Using the formulas (\ref{pl4}) and (\ref{pl5}) we get the twist two-tensor
$F_{\kappa,\kappa'}$ corresponding the classical $r$-matrix (\ref{pl3})
\begin{equation}\label{pl6}
\begin{array}{rcl}
\displaystyle F_{\kappa,\kappa'}\,=\,\lim\limits_{R\to\infty}\,
F\Bigl(\frac{-\imath}{\kappa R},\frac{-\imath}{\kappa'R}\Bigr)
\!\!&:=\!\!&\displaystyle\exp\Bigl({\frac{\imath\kappa}{\kappa'}\sigma_{+}
\otimes(N_{1}-M_{3})e^{-\sigma_{+}}}\Bigr)
\\[10pt]
&&\!\!\!\!\!\!\displaystyle\times\,\exp\Bigl({\frac{\imath}{\kappa}P_{1}
\otimes(N_{1}-M_{3})e^{-\sigma_{+}}}\Bigr)\exp\bigl({\imath N_{2}
\otimes\sigma_{+}}\bigr)~.
\end{array}
\end{equation}
By applying the rescaling of the formulas (\ref{jt17})--(\ref{jt22}) (introducing
dS radius $R$ and performing the limit $R\rightarrow\infty$) we obtain the list of
deformed co-products for all generators of the $D=3$ Poincar\'{e} algebra.
For convenience in what follows we set $\Lambda_{+}:=e^{-\sigma_{+}}-1$,
$L_{+}:=N_{1}-M_{+}$ and $L_{-}:=N_{1}+M_{+}$. The coproducts are given by
\begin{eqnarray}\label{pl7}
\Delta_{\kappa,\kappa'}(\sigma_{+})\!\!&=\!\!&\sigma_{+}\otimes 1+
1\otimes\sigma_{+}~,
\\[10pt]\label{pl8}
\Delta_{\kappa,\kappa'}(P_{1})\!\!&=\!\!&P_{1}\otimes e^{-\sigma_{+}}+
1\otimes P_{1}+\frac{\kappa^2}{\kappa'}\bigl(\Lambda_{+}\otimes\sigma_{+}
e^{-\sigma_{+}}-\sigma_+\otimes\Lambda_{+}\bigr),\phantom{aaaaaaaaaa}
\end{eqnarray}
\begin{equation}\label{pl9}
\begin{array}{rcl}
\Delta_{\kappa,\kappa'}(P_{-})\!\!&=\!\!&
\displaystyle P_{-}\otimes e^{-\sigma_{+}}
+1\otimes P_{-}-\frac{1}{\kappa}\bigl(2P_{1}^{}\otimes P_{1}^{}\,e^{-\sigma_{+}}
+P_{1}^2\otimes\Lambda_{+}e^{-\sigma_{+}}\bigr)
\\[5pt]
&&\displaystyle-\,\frac{2\kappa}{\kappa'}\bigl(P_{1}^{}e^{-\sigma_{+}}\otimes
\sigma_{+}e^{-\sigma_{+}}-\sigma_{+}\otimes P_{1}^{}e^{-\sigma_{+}}
-P_{1}^{}\Lambda_{+}^{}\otimes\sigma_{+}\Lambda_{+}e^{-\sigma_{+}}
\\[5pt]
&&\displaystyle-\,P_{1}^{}\sigma_{+}\otimes\Lambda_{+}e^{-\sigma_{+}}+
\Lambda_{+}\otimes P_{1}^{}e^{-\sigma_{+}}\bigr)-\frac{\kappa^3}{{\kappa'}^2}
\bigl(\Lambda_{+}e^{-\sigma_{+}}\otimes\sigma_{+}^{2}e^{-\sigma_{+}}
\\[10pt]
&&\displaystyle+\,
\sigma_{+}^{2}\otimes\Lambda_{+}e^{-\sigma_{+}}+
\Lambda_{+}^{2}\otimes\sigma_{+}^{2}\Lambda_{+}e^{-\sigma_{+}}-
2\Lambda_{+}\sigma_{+}\otimes\sigma_{+}\Lambda_{+}e^{-\sigma_{+}}\bigr),
\end{array}
\end{equation}
\begin{eqnarray}\label{pl10}
\Delta_{\kappa,\kappa'}(L_{+})\!\!&=\!\!&L_{+}\otimes
e^{\sigma_{+}}+e^{\sigma_{+}}\otimes L_{+}~,
\phantom{aaaaaaaaaaaaaaaaaaaaaaaaaaaaaaa}
\end{eqnarray}
\begin{equation}\label{pl11}
\begin{array}{rcl}
&&\Delta_{\kappa,\kappa'}(N_{2})\,=\,\displaystyle N_{2}\otimes e^{-\sigma_{+}}
+1\otimes N_{2}+\frac{1}{\kappa}\,P_{1}\otimes L_{+}e^{-2\sigma_{+}}
-\frac{\kappa}{\kappa'}\bigl(L_{+}e^{-\sigma_{+}}\otimes\Lambda_{+}
\\[9pt]
&&\qquad\quad\displaystyle-\,L_{+}e^{-2\sigma_{+}}\otimes\sigma_{+}e^{-\sigma_{+}}
-\Lambda_{+}\otimes L_{+}(\sigma_{+}+1)e^{-2\sigma_{+}}+
\sigma_{+}\otimes L_{+}e^{-2\sigma_{+}}\bigl),
\end{array}
\end{equation}
The last coproduct appears very lengthly and we present it only in the
limit $\kappa'\rightarrow\infty$:
\begin{equation}\label{pl12}
\begin{array}{rcl}
\Delta_{\kappa}(L_{-})\!\!&=\!\!&\displaystyle L_{-}\otimes e^{-\sigma_{+}}+
1\otimes L_{-}+\frac{1}{\kappa}\bigl(P_{-}\otimes L_{+}e^{-\sigma_{+}}-
\\[9pt]
&&\displaystyle -\,P_{1}\otimes\{N_{2},e^{-\sigma_{+}}\}
-2N_{2}\otimes P_{1}e^{-\sigma_{+}}-2N_{2}P_{1}\otimes
\Lambda_{+}e^{-\sigma_{+}}\bigr)
\\[5pt]&&\displaystyle-\,\frac{1}{\kappa^2}\bigl(P_{1}\otimes\{P_{1},L_{+}\}
e^{-2\sigma_{+}}+P_{1}^{2}\otimes L_{+}(2\Lambda_{+}+1)e^{-2\sigma_{+}}\bigr).
\end{array}
\end{equation}
Using the formulas (\ref{jt23})--(\ref{jt27}) one can calculate
the antipodes
\begin{eqnarray}\label{pl13}
S_{\kappa,\kappa'}(\sigma_{+})\!\!&=\!\!&-\sigma_{+},\quad
S_{\kappa,\kappa'}(P_{1})=-P_{1}\,e^{\sigma_{+}},
\\[7pt]\label{pl14}
S_{\kappa,\kappa'}(P_{-})\!\!&=\!\!&-P_{-}\,e^{\sigma_{+}}-
\frac{1}{\kappa}\,P_{1}^2\,(e^{2\sigma_{+}}+e^{\sigma_{+}}),
\\[3pt]\label{pl15}
S_{\kappa,\kappa'}(L_{+})\!\!&=\!\!&-L_{+}e^{-2\sigma_{+}},\quad
S_{\kappa,\kappa'}(N_{2})\,=\,-N_{2}\,e^{\sigma_{+}}+\frac{1}{\kappa}\,P_{1}\,L_{+},
\end{eqnarray}
\begin{equation}\label{pl16}
S_{\kappa,\kappa'}(L_{-})\,=\,-L_{-}e^{\sigma_{+}}-\frac{1}{\kappa}\bigl(
P_{-}L_{+}+2N_{2}P_{1}(e^{2\sigma_{+}}\!+e^{\sigma_{+}})\bigr)+
\frac{1}{\kappa^2}\,P_{1}^2L_+(2e^{\sigma_{+}}\!+1).
\end{equation}

\setcounter{equation}{0}
\section{Outlook}
In this paper we have described firstly in detail the Hopf algebra structure of
the two-parameter twisted $D=4$ Lorentz algebras, which induce as well the twist
quantization of the Poincar\'{e} algebra. It should be stressed that contrary
to the cases of twisted Lorentz and Poincar\'{e} symmetries with Abelian twists
which were considered recently in the literature \cite{W,CKNT,ABDMSW,LW,GKMG} our
deformation is generated by the non-Abelian twist. We also performed the dS
contraction limit of the twisted $D=4$ Lorentz algebra and obtained the
deformation of the $D=3$ Poincar\'{e} algebra with the explicit Hopf
algebra structure.

A question which should be further addressed is
the Hopf structure of dual quantum Poincar\'{e} group and the description of
corresponding noncommutative Minkowski space. The quantum Poincar\'{e} groups
generated by the classical Lorentz $r$-matrix (\ref{v11}) can be found on the
list of deformed Poincar\'{e} groups given by Podle\'{s} and Woronowicz \cite{PW}.
Other method providing a quantum algebra of noncommutative Poincar\'{e} group
parameters is to calculate in adjoint $4\times4$ matrix representation the
quantum $R$-matrix and apply the RTT method \cite{FRT}. In our case the
universal $R$-matrix takes the form
\begin{equation}\label{Ou1}
\begin{array}{rcl}
R\;=\;F^{21}(\alpha,\beta)\,F^{-1}(\alpha,\beta)\!\!&=\!\!&\displaystyle
\exp{\Bigl(\frac{\beta}{\alpha^2}\;\varphi\wedge\sigma\Bigr)}\,
\exp{(\sigma\otimes h-\varphi\otimes h')}
\\[10pt] &&\times\;
\displaystyle\exp{(h'\otimes\varphi-h\otimes\sigma)}\,\exp{\Bigl(
\frac{\beta}{\alpha^2}\;\varphi\wedge\sigma\Bigr)}~.
\end{array}
\end{equation}
Using the adjoint matrix representation of Lorentz generators
$(M_{ij})_{\mu\nu}=\delta_{\mu i}\delta_{\nu j}-\delta_{\mu j}\delta_{\nu i}$,
$(M_{0j})_{\mu\nu}=\delta_{\mu 0}\delta_{\nu j}-\delta_{\mu j}\delta_{\nu 0}$
($\mu,\nu=0,1,2,3$; $i,j=1,2,3$) we can obtain from (5.1) a
$16\times16$-dimensional quantum adjoint $R$-matrix for the Lorentz group.
The $R$-matrix (\ref{Ou1}) can be treated also as the universal $R$-matrix for
$D=4$ Poincar\'{e} algebra, and one can obtain easily the $25\times25$-dimensional
quantum $R$-matrix for $D=4$ Poincar\'{e} algebra\footnote{For that purpose one
should insert in (\ref{Ou1}) the $5\times 5$-dimensional realization of  Poincar\'{e}
algebra, with the momenta generators $P_\mu$ described by the matrices having only
the fifth column nonvanishing $(P_\mu=\delta_{\mu 5})$.}.

It is well-known that one can obtain the $D=4$ $\kappa$-deformed Poincar\'{e}
algebra in its standard form (with time variable quantized) if we perform the
quantum AdS contraction of $q$-deformed AdS algebra $U_{q}(\mathfrak{so}(3,2))$ (see \cite{LNRT}).
In Sect.5 we have shown that if we treat the $D=4$ Lorentz algebra as the de-Sitter
algebra one obtains by the suitable quantum dS contraction the generalized light
cone $\kappa$-deformation of the $D=3$ Poincar\'{e} algebra, with two independent
masslike parameters $\kappa$ and $\kappa'$. Therefore one can conclude that quantum
extension of Inonu--Wigner contraction procedure to the twisted de-Sitter (or
anti-de-Sitter) Hopf algebras can be used as the derivation method of twisted
Poincar\'{e} symmetries.

\subsection*{Acknowledgments}
The paper has been supported by KBN grant 1PO3B01828 (A.B., J.L.)
and the grants RFBR-05-01-01086, INTAS-OPEN-03-51-3350 (V.N.T.).
The third author would like to thank Institute for Theoretical
Physics, University of Wroc{\l}aw for hospitality.

\end{document}